\documentclass[11pt]{aastex}

\usepackage{graphicx}

\begin{document}

\title{The Snake - a Reconnecting Coil in a Twisted Magnetic Flux Tube}

\author{Geoffrey V. Bicknell}
\affil{Research School of Astronomy \& Astrophysics and Department of Physics and Theoretical Physics,
Australian National University}
\email{Geoff.Bicknell@anu.edu.au}

\and
\author{Jianke Li\altaffilmark{1}}
\affil{School of Mathematical Sciences and Research School of Astronomy \& Astrophysics, Australian
National University}
\altaffiltext{1}{Current address: Department of Education, Training and Youth Affairs, 14 Mort Street,
Civic, Canberra, Australia}

\slugcomment{Submited to Astrophysical Journal Letters}

\begin{abstract}
 We propose that the curious Galactic Center filament known as ``The Snake'' is a twisted giant
magnetic flux tube, anchored in rotating molecular clouds. The MHD kink instability generates coils in
the tube and subsequent magnetic reconnection  injects relativistic electrons. Electrons diffuse away
from a coil at an energy-dependent rate producing a flat spectral index at large distances from it.
Our fit to the data of \citet{gray95a} shows that the magnetic field
$\sim 0.4 \> \rm mG$ is large compared to the ambient $\sim 7 \mu \> \rm G$ field, indicating that the
flux tube is force-free. If the {\em relative} level of turbulence in the Snake and the general
interstellar medium are similar, then electrons have been diffusing in the Snake for about $3 \times
10^5 \> \rm yr$, comparable to the timescale at which magnetic energy is annihilated in the major kink.
Estimates of the magnetic field in the G359.19-0.05 molecular complex are similar to our estimate of
the magnetic field in the Snake suggesting a strong connection between the physics of the anchoring
molecular regions and the Snake.  We suggest that the physical processes considered here may be
relevant to many of the radio filaments near the Galactic Center. We also suggest further observations
of the Snake and other filaments that would be useful for obtaining further insights into the physics
of these objects. 
\end{abstract}


\section{Introduction}
The fundamental nature of the numerous radio filaments observed near the Galactic Center (see
\citet{larosa00a} for an overview) is unclear. In particular: How are they formed? What are the sources
of the relativistic electrons? Why do they have such large magnetic field strengths? What is the reason
for the unusually flat spectral indices in some of them? The discovery of the Snake
\citep{gray91a}, a 60 pc by 0.4 pc  filament, approximately 150 pc to the West of Sgr A, resulted in
the additional feature that the characteristics of the radio emission are strongly related to a
morphological  ``kink''. Further observations 
\citep{gray95a}, revealed a major and a minor kink, with the radio intensity and spectral index
systematically varying away from the major kink. Various models have been proposed to explain the
spectacular elongated structure of the Snake (see
\citet{gray95a}). However, none of them have been able to successfully explain its major
features. In our view, the Snake is in many respects one of the least complicated of
the filamentary features close to the Galactic Center and a good physical model may
hold the key to understanding the numerous arcs and filaments in the Galactic Center region. 

We propose that the Snake is a magnetic flux tube with both ends anchored in dense rotating material
(molecular clouds and/or associated HII regions). Differential rotation of the tube at either or both
ends produces a monotonically increasing toroidal magnetic field and when
this reaches a critical value ``coils'' or ``loops'' are formed through localized kink instabilities.
Release of the magnetic energy stored in a coil, through magnetic reconnection (and possibly shocks),
is the source of the relativistic electron energy. Energetic electrons diffuse away from each coil at
an energy-dependent rate causing a flattening spectral index.

\section{Details of the model}
\label{s:details}

Let us take 
$R$ to be the radius of the tube, $L_{\rm tube}$  its length  and $B_\phi$ and
$B_z$ the toroidal and axial components of the magnetic flux density, with magnitude $B$, in a
cylindrical
$(r,\phi,z)$ coordinate system coincident with the central axis of the unperturbed tube. As the
end(s) of the tube is(are) rotated, the toroidal field increases and when
$B_\phi/B_z \sim 1$, the kink instability produces a coil  of magnetic flux
with magnetic energy $\Delta E_m \approx  4^{-1} \pi R^3 B^2$
(\citet{alfven50a}, p117). We identify the observed major kink with such a coil; the minor kink may be
another coil at a different stage of development. Numerical simulations, in a solar physics
context \citep{bazdenkov98a,amo95a}, have shown that  the coil's magnetic energy is
annihilated on a time scale of order a few transit Alfv\'en crossing times, $L_{\rm tube}/v_A$, where
$v_A$ is the Alfv\'en speed. Magnetic reconnection and possibly associated shocks in the coil can
provide a source of energy for the acceleration of electrons to relativistic radio-emitting energies
\footnote{The presence of shocks cannot be inferred from the Bazdenkov et al. simulations since they
assumed an incompressible fluid.}. Once accelerated, the electrons diffuse and we model the diffusion
of particles away from the acceleration region by a one-dimensional diffusion equation, assuming a
uniform cross-section for the tube. Taking
$f(p)$ to be the electron phase-space density, $x$ the spatial distance along the flux tube away from
the major kink,
$K(p)$ the momentum-dependent diffusion coefficient, and $C(p)$ the rate per unit volume of momentum
space at which electrons are injected into the coil by the acceleration process, we adopt the
diffusion equation:
\begin{equation}
\frac {\partial f(p,x,t)}{\partial t} - \frac {\partial}{\partial x}
\left( K(p) \frac {\partial f(p,x,t)}{\partial x} \right) = C(p) \delta(x)
\label{e:diffusion}
\end{equation}
The delta function indicates that we are treating
the coil, located at $x=0$, as an infinitesimally small volume with respect to the rest of the tube. 
We also assume that the timescale for dissipation of energy in the coil is greater
than the time over which the electrons diffuse. We discuss this assumption further below. We adopt a
power-law for
$K(p)$, i.e. $K(p) = K_0 (p/p_0)^\beta$, where
$K_0$ is the diffusion coefficient at an electron momentum, $p_0$ of a GeV/c. We also assume that $C(p)
= C_0 (p/p_0)^{-s}$. Integrating equation~(\ref{e:diffusion}) from $x=0^-$ to $x=0^+$ gives the
boundary condition at $x=0$, namely,
\begin{equation}
\frac {\partial f(p,x,t)}{\partial x} \bigg|_{x=0} = -\frac{1}{2} \frac{C(p)}{K(p)}
\end{equation}
We normalize the diffusion equation and the boundary condition by introducing the distance $L=15
\> \rm pc$ between the
major and minor kinks as a fiducial length together with the normalized variables
$\tau = K(p)t/L^2 = (K_0 t / L^2) \left( p/p_0 \right)^\beta$,  
$\xi = {x}/{L}$ and $ g(\xi,\tau) = 2K_0 C_0^{-1} L^{-1} \, \left( p/p_0 \right)^{s+\beta} \, f(p,x,t)
$. In these variables, the diffusion equation and the boundary condition become
\begin{equation}
\frac {\partial g}{\partial \tau} = \frac {\partial^2 g}{\partial \xi^2}  \qquad \hbox{with} \qquad
\frac {\partial g}{\partial \xi}\bigg|_{\xi = 0} = -1  
\end{equation}
for which the solution is
\begin{equation}
g(\xi,\tau) = \left( \frac {4 \tau}{\pi} \right)^{1/2} \, e^{-\xi^2/4 \tau} 
- \xi \left[ 1 - {\rm erf} \left( \frac {\xi}{\sqrt {4 \tau}}\right) \right]
\end{equation}
The number density of particles per unit Lorentz factor, $\gamma$,  is 
\begin{equation}
N (\gamma,\xi,\tau) = N_0 \, \gamma^{-a} 
\, g(\xi,\tau),
\label{e:N}
\end{equation}
where 
$
N_0 = 2\pi (m_e c)^3 (C_0 L/K_0) \gamma_0^{2+a}
$,
$\gamma_0 = (m_e c)^{-1}p_0$ and $a=s+\beta-2$. We use $N(\gamma)$ to
evaluate the angle-averaged synchrotron emissivity, $<j_\nu>$, as a function of frequency, $\nu$, from
\\
\begin{equation}
<j_\nu(\xi)> = (\sqrt 3 e^2/4 c)  \, N_0 \, \nu_B \,  
\left( {2 \nu}/{3 \nu_B} \right)^{(1-a)/2} \, 
\int_{y_1}^{y_2} y^{(a-3)/2} g(\xi,\tau) \bar F(y) \> dy 
\end{equation}
Here, $\nu_B = eB/(2 \pi m_e c)$ and the angle-averaged single electron emissivity\\ 
$\bar F(y) = y \int_y^\infty \sqrt{1-y^2/u^2} \,  K_{5/3}(u) \> du$. Because of the $\gamma$
dependence in $g(\xi,\tau)$, the spectral index near the kink is not
simply $(a-1)/2$ but $(s+\beta/2-3)/2=(a-1)/2-\beta/4$.

For a uniform flux tube, of radius $R$ and angular diameter $\Phi$, imaged with a circular
Gaussian beam with standard deviation, $\sigma$, the angle-averaged flux density per beam is \\
$F_\nu = 4
(2\pi/3)^{1/2} \, \left( \sigma \Phi \right) \left[ I_0(\Phi^2/12\sigma^2) + I_1(\Phi^2/12\sigma^2)
\right]  (<j_\nu> R)$
where $I_0$ and $I_1$ are modified Bessel functions. Using this expression and the expression
for the emissivity, we have performed a least squares fit to the observed
flux densities \citep{gray95a} at 1446 and 4790~MHz, for which,
$\sigma = 4.90^{\prime\prime}$, $R=0.22 \> \rm pc$ and $\Phi =
9.4^{\prime
\prime}$ are appropriate. We restrict the fit to the region of the Snake between galactic latitudes
of $-3^\prime$ and $-17^\prime$ that is clearly related to the major kink; the radio images and the
spectral index plot (\citet{gray95a}, Figure~12) indicate that the minor kink influences the flux
density south of $b=-17^\prime$. The parameters of the fit (see Figure~1) are
$N_0 = 3.5 \times 10^{-5} \rm cm^{-3}$, $B=0.37 \> \rm mG$, $a=2.14$, $\beta = 0.57$ and $\tau_0 =
K_0 t / L^2 = 0.46$. The deviations of the fit from the model are most likely the result of
systematic effects, such as variation in the field strength and width of the tube, rather than
the statistical uncertainty in the data. The magnetic field can be estimated from the data, since it
is related to the frequency index, which varies significantly along the tube. However,
experimentation with the fit shows that the precision of the estimate of $B$ is not much better than
a factor of a few although the result that the magnetic field is much greater than the ambient
interstellar value $\approx 7 \, \mu \rm G$ \citep{gray95a}, is robust. Therefore, the flux tube is
force-free.

If we assume that the injected electron spectrum extends between momenta, $p_1$ and $p_2$, the total
electron energy in the flux tube can be estimated as follows. Let $A$ be the tube's cross-sectional
area, then the rate, $P_{\rm e}$ of relativistic electron energy injection into the tube, is 
$
P_{\rm e} = 4 \pi A c \int_{p_1}^{p_2} p^3 C(p) \, dp =
 4\pi A C_0 cp_0^4 \, (4-s)^{-1} \,
\left[(p_2/p_0)^{4-s} - (p_1/p_0)^{4-s}\right]
$
The total relativistic electron energy in the flux tube is therefore,
\begin{eqnarray}
E_{\rm e,tot} = P_{\rm e} t &=& 2 \pi R^2 m_e c^2 \, \gamma_0^{2-a} 
\, \left( \frac {K_0 t}{L^2} \right) \, \left( \frac{N_0 L} {4-s} \right) \,
\left[(p_2/p_0)^{4-s} - (p_1/p_0)^{4-s}\right] \nonumber \\
&\approx&  6.3 \times 10^{44} \> \left(  {p_2}/{p_0} \right)^{0.43} \> \rm ergs 
\end{eqnarray}
using the expression for $N_0$ (following equation~(\ref{e:N})) and the derived parameters. Time enters
through the dimensionless model parameter, $\tau_0 = K_0 t/L^2$.
For an upper cutoff of 10~GeV, $E_{\rm e,tot} \approx 3.9 \times 10^{45} \> \rm ergs$ and for $100 \>
\rm GeV$, $E_{\rm tot} \approx 1.1 \times 10^{46} \> \rm ergs$. The available
magnetic energy stored in the coil, $E_B \approx 3.5
\times 10^{46} \> \rm ergs$, exceeds the total energy in electrons, as
required for consistency of the model, but not necessarily by a large factor, depending upon the maximum
energy to which electrons are accelerated. This is a consequence of the high value of the derived
magnetic field. These comparisons also suggest that the diffusion of electrons has been taking place
for a time comparable to the total time available to annihilate the coil, consistent with our not
observing the process at a special epoch and consistent with the assumption of continuous injection.

The energy density of electrons, $\epsilon_{\rm e}$, can be derived simply from equation~(\ref{e:N}) for
$N(\gamma)$. Near the coil ($\xi=0$), 
\begin{equation}
\epsilon_{\rm e} = \frac{N_1 \, m_{\rm e} c^2}{-a+\beta/2+2} \, 
\left[ (p_2/p_0)^{-a+\beta/2+2} - (p_1/p_0)^{-a+\beta/2+2} \right] 
\end{equation}
where $N_1 =2 \pi^{-1/2} N_0 \tau_0^{1/2} \gamma_0^{-\beta/2}$.
The energy density is insensitive to the upper cutoff, and for $p_2 =10 \> \rm GeV/c$,
$\epsilon_{\rm e} \approx 6 \times 10^{-11} \> \rm ergs \> cm^{-3}$, an order of magnitude larger
than the energy density of the ISM. 

\section{The spectrum of hydromagnetic turbulence}

It is usually assumed that relativistic  electrons resonantly scatter off a pre-existing level of
hydromagnetic turbulence since resonant Alfv\'en waves are
damped rapidly in the warm interstellar medium by ion-neutral collisions, although this constraint is
more important at higher then GeV energies (e.g.
\citet{melrose82a}). Let the energy density per
unit wave number, $k$, of resonant Alfv\'en waves (either pre-existing or self-excited) be $W(k) =
W_0 \, (k/k_0)^{-\eta}$, where $k_0=eB/cp_0$ corresponds to waves resonating with
1~GeV electrons. Then, the spatial diffusion coefficient for relativistic electrons is
$
K(p) = {4 \eta (\eta+2)}/{9 \pi} \, (c^2/e) \, B^{-1} p 
\left[ {W_{\rm m}}/{k_R W(k_R)} \right]
$
where $W_{\rm m} = B^2/8\pi$ and $k_R = eB/cp$ is the resonant wave number \citep{melrose82a}.
Numerically,
$
K(p) = 1.4 \times 10^{19} \eta (\eta + 2) \, \left( {B}/{\rm mG} \right)^{-1} \,
\left( {W_{\rm m}}/{k_0 W_0} \right) \, \left(  {p} / {p_0} \right)^{2-\eta}
\> \rm cm^2 \> s^{-1}
$. 
Hence, $\beta = 2-\eta$ and $\eta \approx 1.43$ for the parameters of our model. Our value of
$\beta \approx 0.57$ is close to the \citet{ormes83a} value of $0.7$ derived from a cosmic
ray propagation model. However, more recent models adopt a Kolmogorov value,
$\beta=1/3$, combined with the effects of ``minimal reacceleration'' (e.g. \citet{ptuskin99a}).  We
have yet to take account of this possible effect.  In order to constrain
$\beta$ more effectively, one needs to take into account, not only minimal reacceleration, but the
other physical parameters in the problem such as the width and strength of the flux tube and time
variability in the injection process. 

The time since the coil started to inject electrons along the flux tube is of interest for comparison
with other timescales. This is 
$
t = {L^2}/{K_0} \, \tau_0 \approx 3.1 \times 10^5 \, K_{0,26}^{-1} 
\left( {\tau_0}/{0.46} \right) \> \rm yr
$.
For cosmic rays, \citet{ptuskin99a} take $K_0 \sim 10^{28}
\> \rm cm^2 \> s^{-1}$, implying that
$k_0W_0/W_{\rm m} \sim 10^{-6}$. If the same {\em relative} level of turbulence exists in the Snake,
then $K_0 \sim 10^{26} \> \rm cm^2 \, s^{-1}$ since $K_0 \propto B^{-1}$ and $t \sim 3 \times 10^5 \>
\rm yr$.   According to the numerical simulations of comparatively short wound flux tubes
\citep{bazdenkov98a}, a coil disappears ``explosively'' in
$1-3$ Alfv\'en-crossing times, $L_{\rm tube}/v_A$. For the Snake, this is $\approx
1.5-4.5
\times 10^5 \, ({n/10 \> \rm cm^{-3}})^{1/2} \> \rm yr$ based on $L_{\rm tube}
\approx 60 \> \rm pc$ and a number density $n \sim 10 \> \rm cm^{-3}$ for the ambient interstellar
medium \citep{gray95a}. Explosive bursts recur approximately every 5 Alfv\'en times, i.e.
approximately every
$7.5 \times 10^5 \, ({n/10 \> \rm cm^{-3}})^{1/2} \> \rm yr$. Thus, for  $K_0 \sim 10^{26} \> \rm cm^2
\, s^{-1}$, the time over which the electrons have been diffusing is of order the timescales of the
energy releasing process and, as above, we are not required to invoke a special epoch of observation.
This also again justifies our assumption of continuous injection over the diffusion timescale.

\section{The origin of the magnetic field}

We suggest that a twisted magnetic
flux tube with the properties required by our model would arise in the following way: The
magnetic flux tube is  initially anchored at both ends in molecular clouds before they undergo
contraction and initiate star formation. Since molecular clouds condense from the warm interstellar
medium, the cloud magnetic field at the pre-contraction phase exceeds the general ISM value. Hence
such a flux tube emerging into the ISM from a molecular cloud would expand. As one or both of the
clouds contract and rotate, a significant toroidal field is produced and the resulting magnetic
curvature force draws in the flux tube, thereby increasing the poloidal flux density in the entire
flux tube to the molecular cloud value and eventually leading to a force free state which is also
unstable. (Both a force-free configuration and instability require
$B_\phi$ at the boundary of the flux tube to be of order $B_z$.)

There is reasonably good evidence for  the anchoring of the Snake in molecular clouds or associated
HII regions. Observations by
\citet{uchida96a} reveal that the northern end of the Snake intersects an HII region in a CO
``hole'' in the molecular cloud and HII region complex, G359.19-0.05. The supernova remnant observed
in projection against the southern end
\citep{gray95a} provides at least circumstantial evidence for molecular material in that region.
Another possibility is that the Snake could be part of a giant loop, the other end of which is
anchored in another region of the Galactic Center.  The generation of toroidal field is inextricably
linked to the evolution of magnetic field and angular velocity in contracting molecular clouds. Here we
discuss some of the physics involved. However, it will be evident that a complete analysis involves
substantial issues in the physics of contracting molecular clouds that are beyond the scope of this
letter.

The generation of toroidal fields by contracting clouds is related to the radiation of
torsional Alfv\'en waves by the rotating field anchored in the cloud (see \citet{mestel99a}, p 453). Let
$\rho_0$ be the mass density of the background ISM and let $B_z$ be the (uniform) poloidal flux
density of the cloud and ISM. The angular velocity, $\Omega$ and toroidal component, $B_\phi$ of the
field in the tube are governed by:
\begin{equation}
\rho_0 r \frac {\partial \Omega}{\partial t} = \frac{B_z}{4 \pi} 
\frac {\partial B_\phi}{\partial z} \qquad
\frac {\partial B_\phi}{\partial t} = r B_z \frac{\partial \Omega}{\partial z}
\label{e:dbphi_dt}
\end{equation}
The angular velocity in the background medium in which the Alfv\'en speed is $v_A$, satisfies 
\begin{equation}
\frac{\partial^2 \Omega}{\partial t^2} = v_A^2 \frac {\partial^2 \Omega}{\partial z^2}
\label{e:omega}
\end{equation}
As a consequence of these equations, the toroidal field in a tube which is anchored only in the cloud
is
$B_\phi \approx -r(4\pi\rho_0)^{1/2}\Omega_0 \approx - 1.5 \times 10^{-6} (n/10 \> \rm cm^{-3})^{1/2}
(\Omega_0/\rm km \> s^{-1} \> pc^{-1}) \> \rm G$,  where $\Omega_0$ is the angular velocity of the
cloud \citep{mestel99a}. In order to generate $B_\phi \sim 4 \times 10^{-4} \> \rm G$, one requires an
extraordinarily large value of $\Omega \sim 270 \> \rm km \> s^{-1} \> pc^{-1}$.

Consider now a tube anchored, as well, in another
molecular cloud. The boundary conditions on
$\Omega$ are now, $\Omega = \Omega_0$ at $z=0$ and $\Omega = 0$ at $z=L_{\rm tube}$. On a timescale
long compared to the Alfv\'en time, the relevant solution of equation~(\ref{e:omega}) is  $\Omega =
\Omega_0 (1-z/L_{\rm tube})$, implying from the second of equations~(\ref{e:dbphi_dt}), that 
\begin{equation}
\frac {\partial B_\phi}{\partial t} = - B_0 \frac {r \Omega_0}{L_{\rm tube}}
\label{e:bphi}
\end{equation}
Further, if $\Omega_0$ is constant for a time t, then, after the flux tube has been twisted through an
angle $\Delta \phi = \Omega_0 t$ we have $B_\phi = -r \Delta \phi \, B_z / L_{\rm tube}$. This
expression can be derived simply from flux freezing in a twisted flux tube, neglecting the propagation
time of Alfv\'en waves and is used by
\citet{alfven50a}, for example, in  the theory of the instability of twisted flux tubes referred to
above. The purpose of the derivation here is to make a clear connection with the torsional Alfv\'en
waves involved in the standard treatment of magnetic braking. On the basis of this model,
$B_\phi
\sim B_z$ would be attained for the Snake if
$\Omega t
\sim L_{\rm tube}/R \approx 300 (R/0.2 \> \rm pc)^{-1} \> \hbox{radians}$. This would be the case, for
example, for a cloud rotating at an angular velocity of $30 \> \rm km \> s^{-1} \> pc^{-1}$for a
period of $10^7 \> \rm yr$.

However, the above estimates of the toroidal field are somewhat contrived in that they assume that the
flux tube has a constant radius which is clearly not the case in a contracting cloud. Moreover, the
poloidal field changes under the opposing effects of compression and ambipolar diffusion and the
evolution of the field in the external ISM (see above) is also important. However, the main point
is that the anchoring of the magnetic field at {\em both} ends substantially affects the generation of
toroidal field and the requirements on the angular velocity.

A related issue involves that of subcritical or supercritical contraction of the anchoring cloud(s).
In {\em subcritical} contraction, the initial magnetic field supports the cloud
and contraction occurs as a result of ambipolar diffusion of the magnetic flux together with magnetic
braking. The latter prevents the cloud from approaching centrifugal equilibrium
(e.g. \citet{basu95b} and references therein) and the angular velocity remains low, except for
the latest, rapid stages of contraction when the core becomes critical. On the
other hand, in the supercritical case, the magnetic field cannot halt contraction, and if the cloud has
an initial angular velocity, angular momentum conservation causes it to spin up, until it reaches
centrifugal equilibrium. Further contraction is then the result of magnetic braking
\citep{mestel84a}. Therefore, supercritical contraction may offer the best prospect
for twisting of the flux tube. However, this is a complex issue and investigation is deferred for
future work. Note also, that \citet{basu95b} began their simulations with slowly rotating
clouds.

An independent estimate of the poloidal magnetic field is of interest since this is one of the
key parameters in our model and also determines whether the molecular cloud contraction associated
with the Snake is sub- or supercritical. We use parameters derived by
\citet{uchida96a} who estimate that the clouds adjacent to the CO hole have masses
$\sim 5 \times 10^3 M_\odot$ with radii, $R \sim 2 \>
\rm pc$. If we assume that these clouds have formed through compression of spheroidal regions (with
semi-axes $R_0$ and $KR_0$) of the warm interstellar medium of density $n_0 \sim 10 \>
\rm cm^{-3}$ and with an ambient magnetic field of $B_0 \sim 7 \times
10^{-6} \> \rm G$ \citep{gray95a}, then flux conservation implies that
\begin{eqnarray}
B_z &\approx& \frac {B_0 R_0^2}{R^2} = \frac {B_0}{R^2} 
\left[ \frac {3}{4\pi K} \frac {M}{\mu n_0 m_p}\right]^{2/3} 
\nonumber \\
&\approx& 7 \times 10^{-4} \left( \frac {B_0}{7 \times 10^{-6} \rm G} \right) 
\left( \frac {R}{2 \> \rm pc} \right)^{-2} K^{-2/3}
\left( \frac {M}{5 \times 10^3 M_\odot} \right)^{2/3}
\left( \frac {n_0}{10 \> \rm cm^{-3}} \right)^{-2/3} \> \rm G
\end{eqnarray}
Using expressions derived in \citet{mestel99a} (pp. 429 {\em et seq.}), the critical magnetic field 
\begin{equation}
B_{\rm crit} \approx 6 \times 10^{-5} - 1 \times 10^{-4} \left( \frac {M}{5 \times 10^{3} M_\odot}
\right)
\,
\left( \frac {R}{2 \> \rm pc} \right)^{-2} \> \rm G
\end{equation}
with the lowest (highest) value for a spherical (flattened) cloud. These estimates of $B_z$
and $B_{\rm crit}$ raise the prospect that the clouds in the G359.19-0.05 complex are subcritical
unless the original ISM region is highly prolate ($K >> 1$).  However, our main point is that these
indicative estimates of both the actual and critical fields are similar to our estimate of
$4 \times 10^{-4} \> \rm G$ from our model for the Snake. It is therefore possible that the flux
density in the associated  contracting molecular cloud is not greatly enhanced over
its initial value, as a result of ambipolar diffusion. Many of the molecular clouds in the vicinity may
have fields of this magnitude, but it is not until they are twisted and ``lit up'' by reconnection
induced by an instability that they become observable. Interestingly the \citet{basu95b}
simulations show that the magnetic field outside the supercritical core is indeed time independent.

\section{Discussion}

High resolution images of the Snake near the major kink (\citet{gray95a}, Figures 10 and 11) show
that the Snake appears to be
split in two, similar to flux tubes in the late stages of the
\citet{bazdenkov98a} simulations. This provides additional support for our proposal. As well as being
directly applicable to the Snake, our model may open up a number of appealing possibilities for the
dynamics of other magnetic filaments in the interstellar medium near the Galactic Center. It is
feasible, in an environment with such a large density of molecular clouds, that rotation
of  magnetic flux tubes threading their cores would lead to  kink instabilities and reconnection. The
flattening spectral index of the Snake depends upon the differential diffusion of energetic particles
and this is such a well known phenomenon in the regular ISM that its role in this model is
unremarkable.  The index of the momentum dependence of the diffusion parameter that we have inferred
for the Snake is in the range of values used in models of cosmic ray propagation. More detailed models
raise the prospect of a better determination of this index as well as the other parameters in this
model. In particular, a more precise value of the  index of the momentum dependence of the particle
creation rate, which our current estimates place near the value associated with strong shocks, would
be of interest. The type of data that is essential for more detailed modeling include (1) High
resolution images showing more clearly the structure of the flux tube over its entire length. (2) Flux
densities at different frequencies, enabling one to better constrain the energy dependence of the
diffusion parameter and the magnetic field.

It remains to be shown that flux tubes in the Galactic Center can be wound up to such an extent that
they produce a significant toroidal field. If an anchoring
cloud acquires an angular velocity which persists for long enough that it rotates through 300 radians
($\sim 50$ revolutions), that would be sufficient. Another possibility is that the required winding
could be produced in a contracting cloud, although this idea requires further
investigation. In our independent estimates of the magnetic field in the Snake and in the
related issue of subcritical or supercritical contraction we have concentrated exclusively on the
molecular complex at the northern end of the Snake. Currently, however, there is little known about
the southern end and the masses and rotation rates of molecular clouds in that region will certainly be
of interest. 

\acknowledgments We thank Dr. R. Protheroe for helpful information on cosmic ray propagation, Prof. L.
Mestel for advice concerning the physics of molecular clouds and the referee for constructive comments
on the original version of this paper.


\newpage

\begin{center}
\bf Figure Caption
\end{center}

\noindent
Figure 1: The model fit to the 1446~MHz (filled squares) and 4790~MHz (open circles) flux densities
(Gray et al. 1995) for the Snake. The model fit only applies to the region associated with the 
major kink.

\newpage

\begin{figure}
\centering \leavevmode
\includegraphics[width=7.5in]{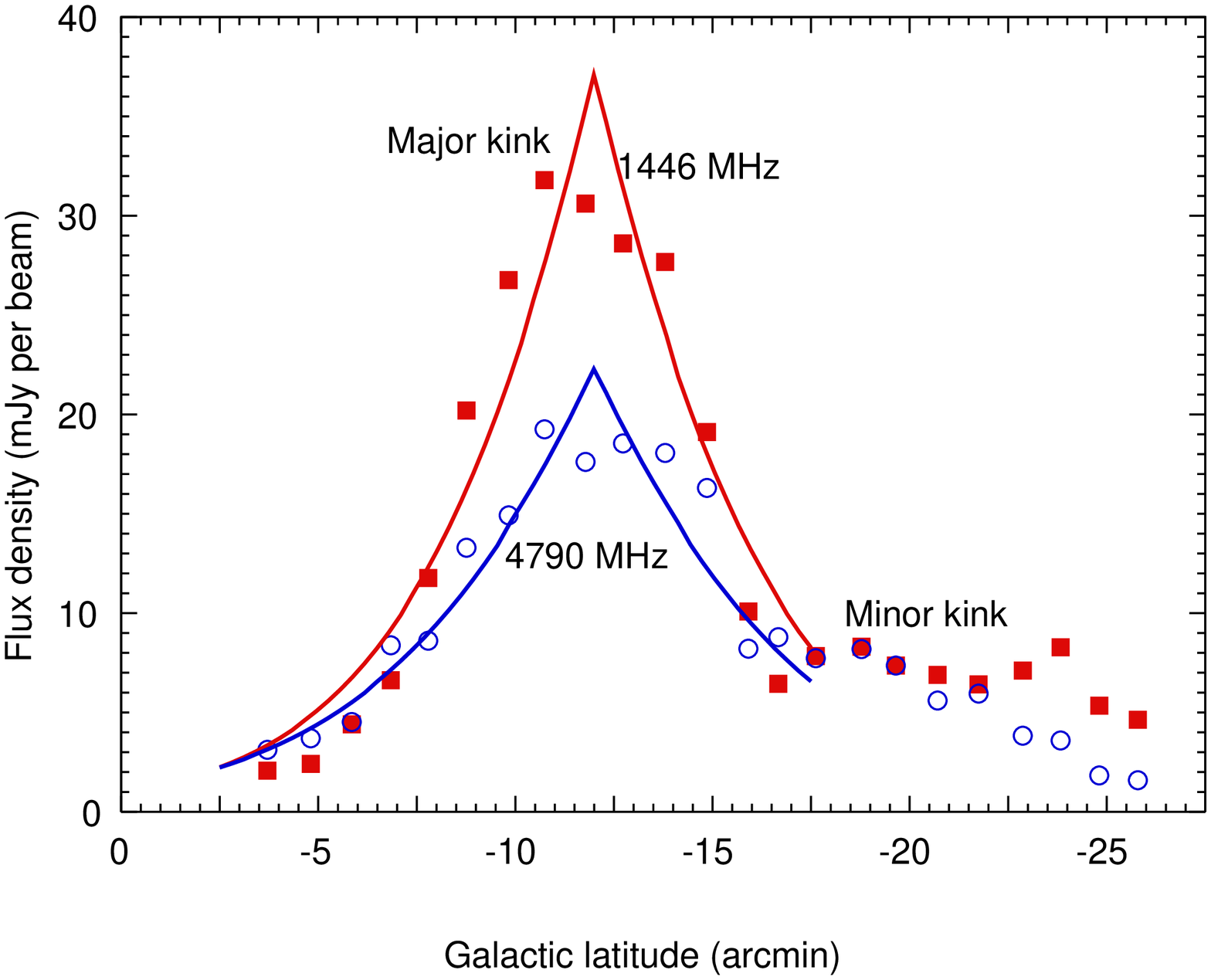}
\caption{}
\label{f:model}
\end{figure}

\end{document}